\documentclass[]{aa}  

\usepackage{graphicx}
\usepackage{txfonts}
\usepackage{hyperref}
\renewcommand{\d}{\mathrm{d}}
\renewcommand{\bf}{}

\begin{document}

   \title{New closed analytical solutions for geometrically thick fluid tori around black holes}
\subtitle{Numerical evolution and the onset of the magneto-rotational instability}

   \author{V. Witzany\thanks{vojtech.witzany@zarm.uni-bremen.de}
          \inst{1}
          \and
          P. Jefremov\thanks{paul.jefremow@zarm.uni-bremen.de}\inst{1}
          }

   \institute{Center of Applied Space Technology and Microgravity (ZARM), Universität Bremen, Am Falturm 2, 28359 Bremen, Germany
             }

   \date{Received 25. 11. 2017; accepted 26. 02. 2018}

 
  \abstract
   {When a black hole is accreting well below the Eddington rate, a geometrically thick, radiatively inefficient state of the accretion disk is established. There is a limited number of closed-form physical solutions for geometrically thick (non-selfgravitating) toroidal equilibria of perfect fluids orbiting a spinning black hole, and these are predominantly used as initial conditions for simulations of accretion in the aforementioned mode. However, different initial configurations might lead to different results and thus observational predictions drawn from such simulations.}
   {We expand the known equilibria by a number of closed multiparametric solutions with various possibilities of rotation curves and geometric shapes. Then, we ask whether choosing these as initial conditions influences the onset of accretion and the asymptotic state of the disk.}
   {We investigate a set of examples from the derived solutions in detail; we analytically estimate the growth of the magneto-rotational instability (MRI) from their rotation curves  and evolve the analytically obtained tori using the 2D magneto-hydrodynamical code HARM. Properties of the evolutions are then studied through the mass, energy, and angular-momentum accretion rates.}
   {The rotation curve has a decisive role in the numerical onset of accretion in accordance with our analytical MRI estimates: In the first few orbital periods, the average accretion rate is linearly proportional to the initial MRI rate in the toroids. The final state obtained from any initial condition within the studied class after an evolution of $\gtrsim 10$ orbital periods is mostly qualitatively identical and the quantitative properties vary within a single order of magnitude. The average values of the energy of the accreted fluid have an irregular dependency on initial data, and in some cases fluid with energies many times its rest mass is systematically accreted.}
   {}


   \maketitle
   

\section{Introduction}

{\bf Accretion of matter onto black holes can lead to the release of large amounts of radiation and is responsible for some of the most energetic phenomena in the Universe such as X-ray binaries or active galactic nuclei. When the accretion rate on a black hole is sufficiently high while radial convection of internal energy sufficiently low}, the accretion disk surrounding the black hole reaches a state which is well described by the so-called thin disk model \citep{shakurasunyaev, novikovthorne,abramowiczlivrev}. However, when the accretion rate drops well below the so-called Eddington accretion rate $\dot{M}_\mathrm{Edd} = 1.39 \times 10^{18} (M/M_\odot) \mathrm{g\,s}^{-1}$, as is believed to be for example the case for the black hole in our Galactic center, the accretion disk enters a rather different mode which is characterized as geometrically thick, optically thin, ``hot'' in the sense that thermal energies are non-negligible as compared to gravitational binding energies, and radiatively inefficient. This also implies that the energy losses from the disk are dominated by either direct advection through the horizon or outflows \citep[see][and references therein]{yuannarayan}.

Amongst other things, this means that the radiatively inefficient disk can be, to a first approximation, modeled as a magneto-hydrodynamic (MHD) fluid without radiation back-reaction \citep[see e.g.][]{dexter2009}. (A more accurate model, however, requires the inclusion of radiative cooling and heating, and the partially two-temperature nature of the fluid in this accretion mode, see e.g. \citet{ryan2017} and references therein.) Consequently, a number of global general-relativistic MHD studies were conducted showing that the MRI-driven turbulence \citep{balbus1991,balbus1998} provides the angular momentum transport in the disk, and many features predicted by analytical or semi-analytical models emerge with only little additional ad-hoc input \citep[see e.g.][]{devilliers2003,mckinney2004,narayan2012}. 

For simulations of this type of accretion disks one usually uses initial conditions which are in a smooth equilibrium state. If such equilibria are to be given completely in closed form, then they are predominantly either the fluid tori with constant specific angular momentum $\ell \equiv -u_\varphi/u_t$ of \citet{kozlowski1978,abramowicz1978}, or with constant $\ell^* \equiv u_\varphi u^t$ of \citet{fish-mon}.

However, the angular momentum or angular velocity distributions play a decisive role in the onset of various instabilities. The MRI growth rate is proportional to the angular velocity gradient, and another notable instability, the centrifugal or Rayleigh-Taylor\footnote{In many contexts the ``Rayleigh-Taylor instability'' is used exclusively for the instability of interfaces of fluids with different densities. Here, we use the word as also applying to the similar effect in continuous media \citep[see e.g.][]{gourgouliatos2017}.} instability, has a growth rate proportional to the gradient of angular momentum. 

Another case where the angular momentum distribution {\bf was formerly believed} to make a crucial difference is the gravitational runaway instability as studied e.g. by \citet{fontdaigneI,fontdaigneII}. {\bf In the particular model chosen by the authors of the aforementioned study, the mass parameter of the black hole space-time was allowed to dynamically grow by the amount of accreted matter, leading possibly to an accretion runaway. Then, it turned out that if specific angular momentum grew with radius, the runaway instability was suppressed, whereas for the $\ell = \mathrm{const.}$ tori it caused the accretion disk to be fully accreted on orbital time-scales. This instability was later shown to disappear when full self-gravity of the torus is accounted for \citep{montero2010,rezzolla2010,mewes2016}. However, the study of \citet{fontdaigneI,fontdaigneII} demonstrates that the precise shape of the angular momentum distribution can often have a critical influence on the evolution obtained within a given model.}

Is it possible that we are missing new dynamical effects or different states of the accretion disks in our simulations because we are restricting ourselves only to a limited number of initial conditions? To answer this question, a variety of closed form solutions corresponding to different gradients of either $\ell$ or $\Omega$ in the initial fluid configurations would be useful.

We present such solutions in this paper in Section \ref{sec:sol}, and demonstrate how these lead, at least in principle, to differences in numerical simulation results in Section \ref{sec:sim} by implementing and evolving our solutions in the HARM 2D code \citep{HARM,HARM2}. A reader looking for an explicit recipe for the construction of the tori will find all the needed formulas in Appendix \ref{app:construction}. Of interest is also the essentially unknown fact that the {\em fully general} solutions for the toroidal equilibria in static space-times (e.g. in the Schwarzschild space-time) are expressible in closed form, which we briefly describe in Subsection \ref{subsec:stat}.


\section{Analytical solutions for fluid tori near black holes} \label{sec:sol}
We use the $G=c=1$ geometrized units and the $-+++$ signature of the space-time metric. {\bf A comma before an index defines a partial derivative with respect to the appropriate coordinate, whereas a preceding semi-colon a covariant derivative with respect to the coordinate.}

\subsection{Basic equations}

The plasma orbiting the black hole is modeled by ideal MHD where molecular dissipation, electric resistivity, self-gravitation, and radiative or non-equilibrium effects are neglected in the dynamics. As a result, the system is governed by the set of equations \citep[cf.][]{anile1989}
\begin{align}
&\left(T^{\mu}_{\;\; \nu}\right)_{;\mu} = \left(T^{\mu}_{\;\; \nu\mathrm{(m)}} + T^{\mu}_{\;\; \nu\mathrm{(f)}} \right)_{;\mu} = 0 \,, \label{eq:Euler}\\
&(\mu u^\mu)_{;\mu} = 0 \,, \label{eq:pc}\\
&(u^\mu b^\nu - u^\nu b^\mu)_{;\nu} = 0\,, \label{eq:faraday}\\
&T^{\mu\nu}_\mathrm{(m)} \equiv b^2 u^\mu u^\nu + \frac{b^2}{2} g^{\mu \nu} -b^\mu b^\nu \,, \\
&T^{\mu\nu}_\mathrm{(f)} \equiv w u^\mu u^\nu + P g^{\mu \nu} \,,
\end{align}
where $u^\mu$ is the fluid four-velocity, $\mu$ its rest-mass density, $w$ its enthalpy density, and $P$ its pressure. The vector $b^\mu$ is the magnetic field vector in the rest frame of the fluid. 

We place the magneto-fluid into an axially symmetric and stationary space-time with the metric
\begin{equation}
\d s^2 = g_{tt} \d t^2  + 2 g_{t \varphi} \d t \d \varphi + g_{\varphi \varphi} \d \varphi^2 + g_{rr} \d r^2 + g_{\vartheta \vartheta} \d \vartheta^2 \,.
\end{equation}
Some useful symbols we will use are the rotation frequency of zero-angular-momentum observers (ZAMOs), $\omega \equiv g^{t\varphi}/g^{tt} = - g_{t\varphi}/g_{\varphi \varphi}$; the potential whose gradient generates the acceleration of ZAMOs, $\Phi_\mathrm{(Z)} \equiv -\ln(-g^{tt})/2$; minus the determinant of the $t$-$\varphi$ part of the metric, $\rho^2 \equiv g_{t\varphi}^2 - g_{tt} g_{\varphi \varphi}$; and a radius-like quantity $\mathcal{R} \equiv \sqrt{-g^{tt} g_{\varphi \varphi}}$. 

Now let us assume that all the properties of the magneto-fluid are axisymmetric and stationary, the flow is purely circular, $u^r=u^\vartheta=0$, and that the magnetic field is purely toroidal, $b^r=b^\vartheta = 0$. Then the Faraday equation (\ref{eq:faraday}) and the particle-conservation equation (\ref{eq:pc}) are trivially fulfilled and we need to solve only the momentum balance equation (relativistic Euler equation) (\ref{eq:Euler}). 

The magnetic part of this equation can under the given assumptions be simplified as
\begin{equation}
\left(T^{\mu}_{\;\; \nu\mathrm{(m)}}\right)_{;\mu} = \frac{(b^2 \rho^2)_{,\nu}}{2 \rho^2} \label{eq:simplemag}
\end{equation}
This expression was first derived by \citet{komissarov}, but we provide, {\bf in our view, a more direct and clear} rederivation in Appendix \ref{app}.

There are many different ways how to rewrite the fluid part of the Euler equation, one of them is the form which can be attributed to \citet{fish-mon}
\begin{equation}
\begin{split}
\left(T^{\mu}_{\;\; \nu\mathrm{(f)}}\right)_{;\mu} = &w \left[-\ln(\mathcal{R})_{,\nu} V^2 + \omega_{,\nu} \mathcal{R} V \sqrt{1+V^2}  +\Phi_{\mathrm{(Z)},\nu}\right] \\
													 &+ P_{,\nu}\,, \label{eq:fmform}
\end{split}
\end{equation}
where $V \equiv u_\varphi/\sqrt{g^{\varphi \varphi}}$ is the linear velocity of the fluid as measured in the ZAMO frame. 

{\bf However, one of the best known forms of the fluid part of the Euler equation for circular equilibria is due to }\citet{kozlowski1978,abramowicz1978} and reads
\begin{equation}
\begin{split}
\left(T^{\mu}_{\;\; \nu\mathrm{(f)}}\right)_{;\mu} = &w \left[ -\ln(u^t)_{,\nu} + \frac{\ell}{1 - \Omega \ell} \Omega_{,\nu} \right] + P_{,\nu}\,, \label{eq:pdform}
\end{split}
\end{equation}
where $\ell \equiv -u_\varphi/u_t$ and $\Omega = u^\varphi/u^t$. In all of the equations (\ref{eq:simplemag}), (\ref{eq:fmform}) and (\ref{eq:pdform}) the non-trivial components are of course only those with $\nu=r,\vartheta$. We provide brief descriptions of the derivations of the expressions (\ref{eq:fmform}) and (\ref{eq:pdform}) in Appendix \ref{app}.


\subsection{Static space-times} \label{subsec:stat}

In static space-times ($g_{t\varphi} = 0$) the full Euler equation using (\ref{eq:simplemag}) and (\ref{eq:fmform}) simplifies as
\begin{equation}
-\ln(\mathcal{R})_{,\nu} V^2 = -\frac{P_{,\nu}}{w} - \frac{\tilde{P}_{,\nu}}{\tilde{w}} -\Phi_{\mathrm{(Z)},\nu} \label{eq:stat}\,,
\end{equation}
where we have introduced the notation $\tilde{P} \equiv b^2 \rho^2,\, \tilde{w} \equiv 2 w \rho^2$.

Let us now further assume that the configuration of the fluid fulfills the barotropic\footnote{The fluid fulfilling the condition $P=P(w)$ is called barotropic by \citet{kozlowski1978,abramowicz1978}, while other authors would rather call a fluid where the conditions $P=P(\mu)$ is fulfilled barotropic \citep[see e.g.][]{tooper1965}. Both conditions coincide in flows with constant entropy per particle and in the Newtonian limit.} condition of coincident surfaces of constant pressure and enthalpy $P=P(w)$ and the ``magnetotropic'' condition of coincident constant $\tilde{P}$ and $\tilde{w}$, $\tilde{P}= \tilde{P}(\tilde{w})$ \citep[see][]{komissarov}. Then the right-hand side of (\ref{eq:stat}) is a coordinate gradient, and the same must be true also for the left-hand side. Thus we conclude that under the given conditions we necessarily have $V = V(\mathcal{R})$ and the general solution of the Euler equation reads

\begin{equation}
W(w(r,\vartheta)) + \tilde{W}(\tilde{w}(r,\vartheta)) = -\Phi_\mathrm{(c)}(\mathcal{R}(r,\vartheta)) - \Phi_\mathrm{(Z)}(r,\vartheta)\,.
\label{eq:statsol}
\end{equation}
{\bf where we have defined the effective potentials}
\begin{align}
& W(w) \equiv \int \frac{P'(w)}{w} \d w \,, \label{eq:W}\\
&\tilde{W}(\tilde{w}) \equiv \int \frac{\tilde{P}'(\tilde{w})}{\tilde{w}} \d \tilde{w} \,, \label{eq:Wt}\\
& \Phi_\mathrm{(c)}(\mathcal{R}) \equiv -\int \frac{V(\mathcal{R})^2}{\mathcal{R}} \d \mathcal{R}\,, \label{eq:Phi} 
\end{align}
{\bf and where the potentials are determined only up to integration constants.} The functions $W$ and $\tilde{W}$ can be understood as thermodynamic and magneto-thermodynamic effective potentials respectively, $\Phi_\mathrm{c}$ has clearly the meaning of a centrifugal potential, and $\Phi_\mathrm{(Z)}$ is, in the currently discussed case of static space-times, the gravitational potential experienced by static observers. 

General solutions for the toroidal equilibria can then be constructed by
\begin{enumerate}
\item prescribing arbitrary  {\bf distributions} $P(w), \tilde{P}(\tilde{w})$ \citep[{\bf e.g. as an ``enthalpic polytrope''} $P = K w^\gamma,\, \tilde{P} = \tilde{K} \tilde{w}^\kappa $, see][]{komissarov,wielgus2015} and an equally arbitrary rotation profile $V(\mathcal{R})$;
\item analytically integrating the potentials {\bf from} (\ref{eq:W}), (\ref{eq:Wt}) and (\ref{eq:Phi});
\item solving  (\ref{eq:statsol}) for $w(r,\vartheta)$;
\item using $b^2 = \tilde{P}(2 \rho^2 w(r,\vartheta))/\rho^2$ and the relation $b^\mu u_\mu = 0$ to determine $b^t,\, b^\varphi$;
\item postulating an equation of state such as the ideal-gas $w = \mu + 5\mu k_\mathrm{B} T/(2m)$, where $T$ is the temperature and $m$ the particle mass, to derive the density and entropy field from fundamental thermodynamic relations and the function $P(w)$.
\end{enumerate}

The results presented in this Subsection are probably not original, even though it is hard to find their explicit statement in the literature. They can be, however, seen as easily obtainable consequences of the results of either \citet{vonZeip} or \citet{fish-mon} when restricted to static space-times. {\bf 
	
Nonetheless, we find it important to state these results explicitly because there is a number of studies that construct tori in static space-times and use somewhat complicated numerical methods tailored for stationary space-times, even though the above-stated general closed solution is available \citep[e.g.][]{fontdaigneII,narayan2012, penna2013}}.


\subsection{Stationary space-times}
We have just seen that the form of the Euler equation coming from (\ref{eq:fmform}) can be used to construct the entirely general barotropic and magnetotropic solutions in static space-times. Furthermore, \citet{fish-mon} used this form even in the case of stationary space-times to obtain equilibria with constant $\ell^*$. However, we found it much more fruitful to use (\ref{eq:pdform}) in the general $g_{t\varphi} \neq 0$ stationary space-times. 

We obtain the Euler equation under the assumption of barotropicity and magnetotropicity as
\begin{equation}
-\ln(u^t)_{,\nu} + \frac{\ell}{1 - \Omega \ell} \Omega_{,\nu} = - W_{,\nu} - \tilde{W}_{,\nu} \label{eq:PDeuler}
\end{equation}
Once again, this equation quickly leads to an integrability requirement that either $\Omega_{,\mu} = 0$ or $\ell = \ell(\Omega)$. This result is known as the relativistic von Zeipel theorem \citep{vonZeip}. 

However, finding closed analytical solutions is more complicated than in the static case; if we postulate the function $\ell(\Omega)$, we obtain 
\begin{equation}
W + \tilde{W} = - L(\Omega) -\frac{1}{2} \ln (-g_{tt} -2 g_{t\varphi} \Omega - g_{\varphi \varphi} \Omega^2 )\,,
\end{equation}
where we have used the fact that $u^t = (-g_{tt} - 2 g_{t \varphi} \Omega - g_{\varphi \varphi} \Omega^2 )^{-1/2}$ and we define $L(\Omega)$ analogously to the potentials (\ref{eq:W} - \ref{eq:Phi}),
\begin{equation}
L(\Omega) \equiv \int \frac{\ell(\Omega)}{1 - \Omega \ell(\Omega)} \d \Omega\,.
\label{BigL}
\end{equation}
In other words, the thermodynamic and magneto-thermodynamic potentials are specified by this procedure {\bf not only as functions of coordinates but also of an as-of-yet unspecified function $\Omega = \Omega(r,\vartheta)$}. 

To obtain $\Omega(r,\vartheta)$ from the postulated $\ell(\Omega)$ and thus find the full explicit solution, one has to use the relation
\begin{equation}
\Omega= \frac{u^{\varphi}}{u^t}=\frac{\ell g^{\varphi \varphi}-g^{t \varphi}}{\ell g^{t \varphi}-g^{t t}}\,, \label{eq:Omzlomek}
\end{equation}
which leads to the equation for $\Omega$
\begin{equation}
(\ell g^{t \varphi}-g^{t t}) \Omega -\ell g^{\varphi \varphi}+g^{t \varphi}=0\,. \label{eq:Omkvadra}
\end{equation}
In general, this equation has to be solved numerically, see examples for non-magnetized tori in \citet{chakrabarti1985,fontdaigneII,qian2009,penna2013}, and also more recent examples for magnetized tori in \citet{wielgus2015,gimeno2017}. Nevertheless, if we instead want to obtain $\Omega(r,\vartheta)$ as a closed analytical expression, we must impose some constraints on the form of $\ell(\Omega)$. For instance, if $\Omega$ is to be a root of a polynomial equation of order $n$, it is easy to see that necessarily $\ell = P^{(n-1)}(\Omega)/P^{(n-1)}(\Omega)$, where $P^{(k)}(x)$ are some polynomials of order $k$. 

We choose here to focus on the general form of $\ell(\Omega)$ which leads to the expression for $\Omega$ as a root of a quadratic equation (n=2)
\begin{equation}
\ell = \frac{\ell_0 + \lambda \Omega}{1 + \kappa \ell_0 \Omega}\,, \label{eq:ellstar}
\end{equation}
where $\ell_0, \kappa, \lambda$ are some constants.

We choose this particular form of the parametrization of $\ell(\Omega)$ because the case $\kappa = 0, \lambda=0$ corresponds to the well-known $\ell = \mathrm{const.}= \ell_0$ Polish donuts of \citet{kozlowski1978,abramowicz1978}, and, on the other hand, $\kappa=1,\lambda=0$ corresponds to $\ell^* \equiv u_\varphi u^t = \mathrm{const.} = \ell_0$ of \citet{fish-mon}. The other choices of $\kappa,\lambda$, however, represent a continuous class connecting and extending these two particular solution families. 

The expressions for $\Omega$ and the thermodynamic potential as functions of coordinates corresponding to equation (\ref{eq:ellstar}) are given in Appendix \ref{app:construction}; examples of the obtained rotation curves and density profiles are discussed in the next Section.  In particular, it is shown that in Kerr space-time the choice $\lambda=0, \ell_0 >0$ and a small $\kappa>0$ generically leads to $\ell$ growing with radius, whereas $\kappa<0$ to an angular momentum falling of with radius.

{\bf Even more specifically, we will see in the next Section that $\kappa < 0$ leads to very thick tori with strongly non-Keplerian rotation profiles. However, since the solutions themselves do not have any direct physical meaning, and since they are used only as initial conditions for an MHD evolution that completely changes their character, we do not exclude any choice of parameters, save perhaps for those leading to outright pathological tori. In return, this allows us to potentially discover new states and behaviors of accretion disks.}

	\begin{figure}
    	\centering
		\includegraphics[width=0.49\textwidth]{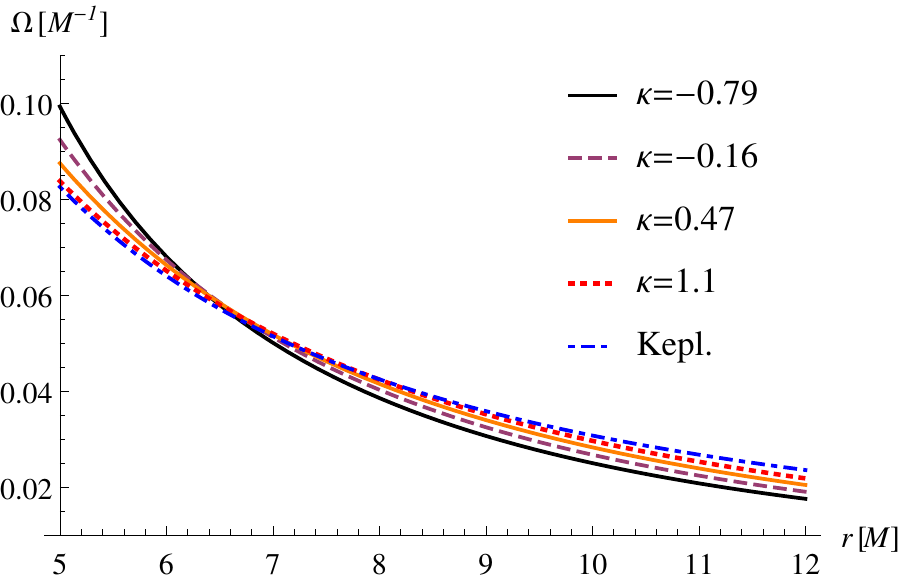}
        \caption{Profile of the angular velocity $\Omega$ inside the tori along the equatorial plane for different values
        of the parameter $\kappa$. In this and all the following graphs we assume
        $\lambda=0$ and $\ell_0$ is determined by the constraint that the torus has its inner and outer edge at $r=5M$ and $r=12M$ respectively.} \label{fig:Om}
	\end{figure}
    \begin{figure}
    	\centering
		\includegraphics[width=0.49\textwidth]{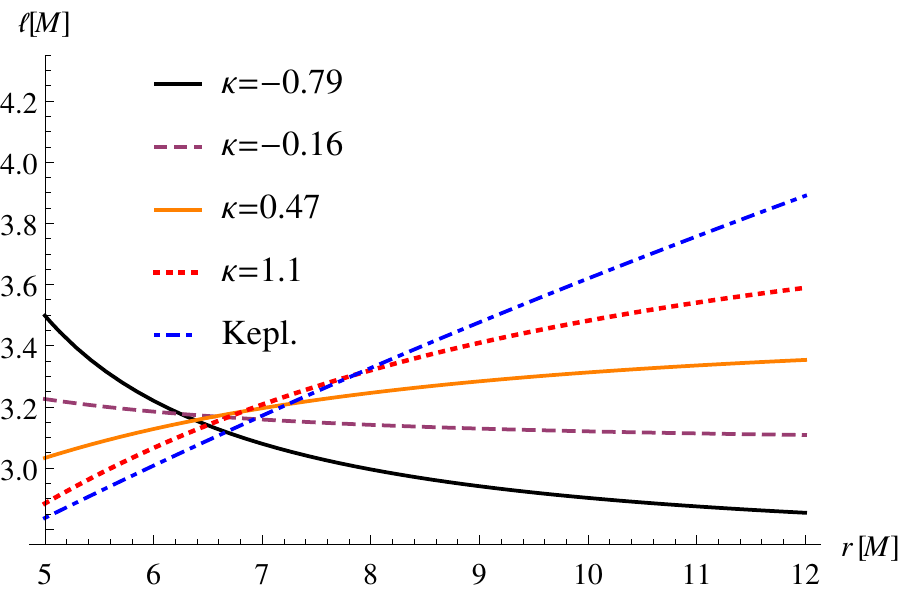}
        \caption{Profile of specific angular momentum $\ell$ inside the tori along the equatorial plane
        for different values of the parameter $\kappa$.} \label{fig:ell}
	\end{figure}

	\begin{figure}
    	\centering
		\includegraphics[width=0.49\textwidth]{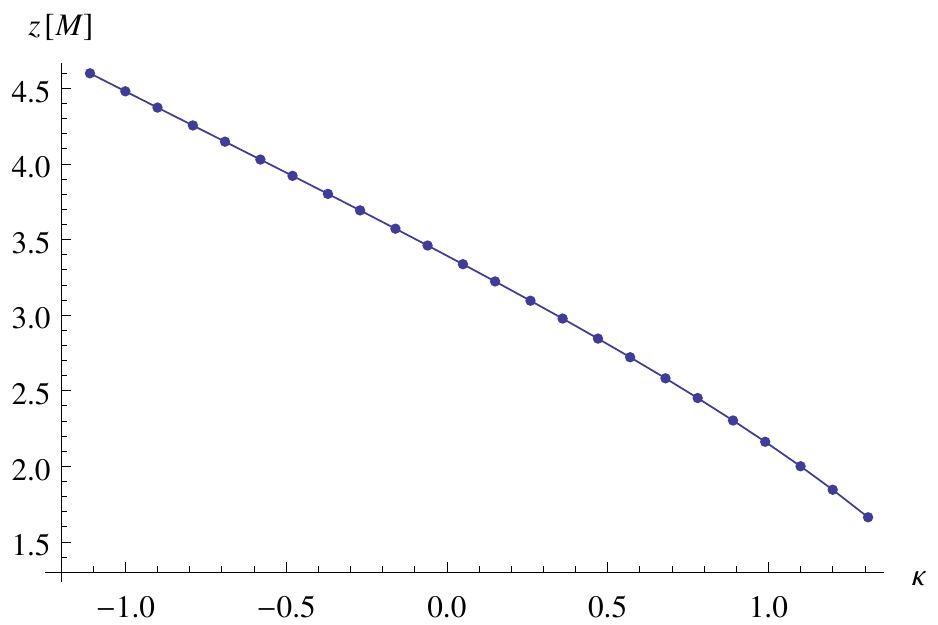}
        \caption{The vertical extent $z= r\cos\vartheta$ of the tori as a function of the parameter $\kappa$.} \label{fig:z}
	\end{figure}
    

		\begin{figure*}
    	\centering
        \includegraphics[width=0.99\textwidth]{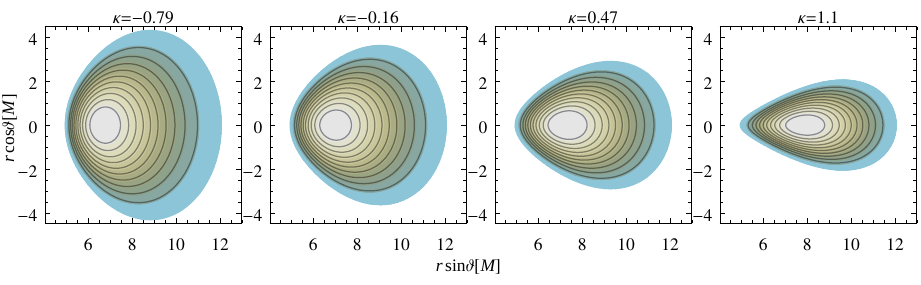}
        \caption{Meridional sections of the density of the tori for different values of $\kappa$. The progression goes from geometrically very thick tori with large pressure gradients at their inner edge up to thinner toroids with a cusp (vanishing pressure gradient) gradually forming at their inner radius.} \label{fig:shapes}
	\end{figure*}

	\begin{figure}
    	\centering
		\includegraphics[width=0.49\textwidth]{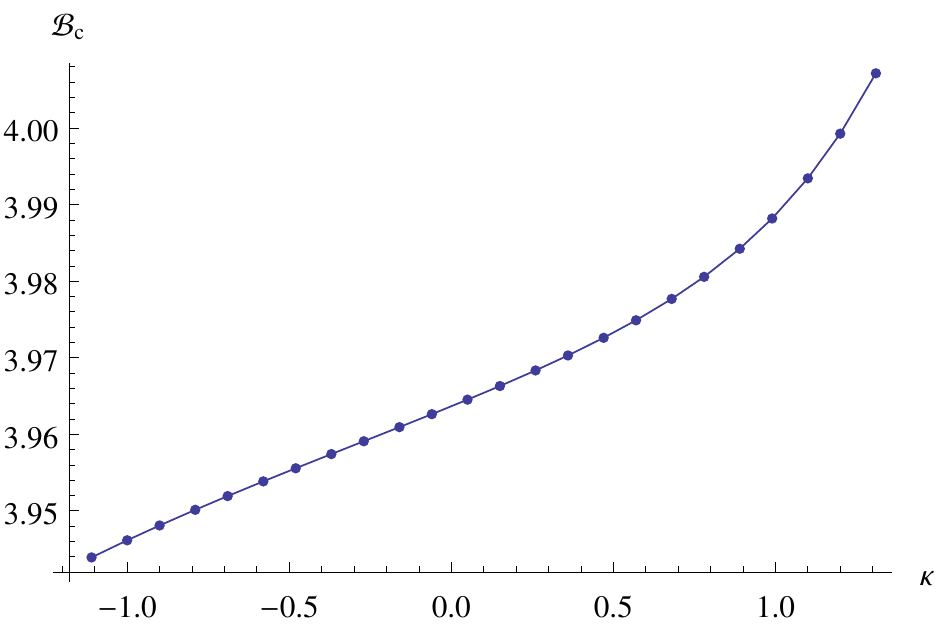}
        \caption{The relativistic Bernoulli  parameter $\mathcal{B} \equiv - h u_t$ at the pressure maxima of the tori ($\mathcal{B}>1$ corresponds to matter with enough energy to escape to infinity) as a function of the parameter $\kappa$. In our case the thermodynamic properties are chosen such that $h=4.25$ at the {\bf pressure maxima} of the toroids, so the behavior of $\mathcal{B}_\mathrm{c}$ is due to the fact that the maxima are getting farther from the black hole with growing $\kappa$.} \label{fig:bernoulli}
	\end{figure}

	\begin{figure}
    	\centering
		\includegraphics[width=0.49\textwidth]{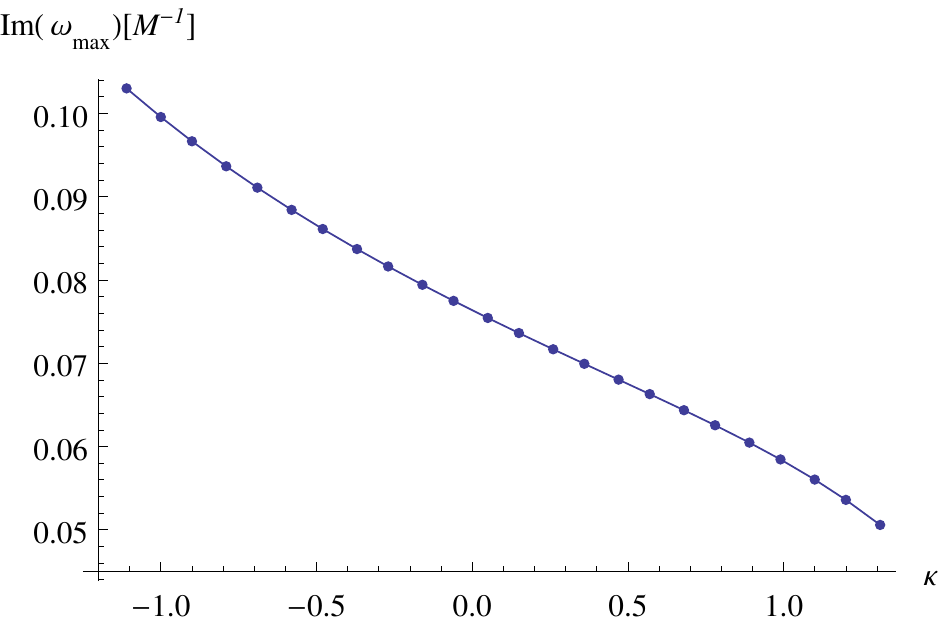}
        \caption{{\bf The growth rates of the fastest-growing MRI mode $\mathrm{Im}(\omega_\mathrm{max})$ evaluated at the position of the pressure maxima of the tori plotted as a function of the parameter $\kappa$.}}
        \label{fig:MRIestim}
	\end{figure}

\begin{figure}
    	\centering
		\includegraphics[width=0.49\textwidth]{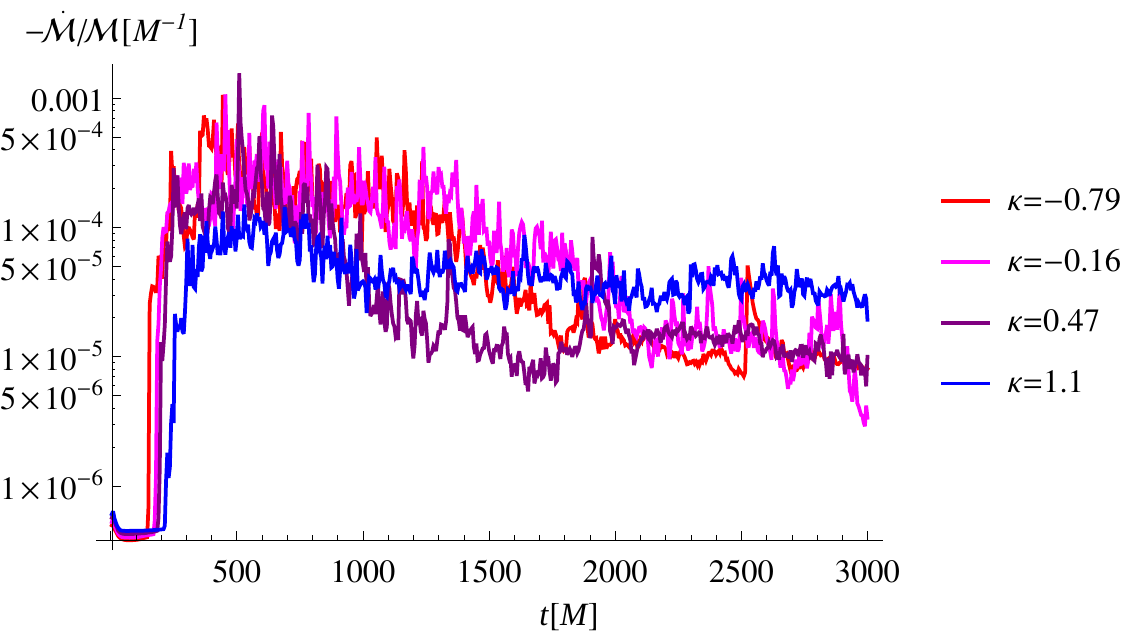}
        \caption{Some examples of the numerically obtained mass accretion rate $\dot{\mathcal{M}}/\mathcal{M}$ as a function of time. The height of the first peak of the accretion rate is correlated with $\kappa$, but the late evolutions are harder to relate to the parameters of the initial conditions. (Color online.)} \label{fig:accexamples}
	\end{figure}


\begin{figure*}
    	\centering
\raisebox{-0.5\height}{\includegraphics[width=0.44\textwidth]{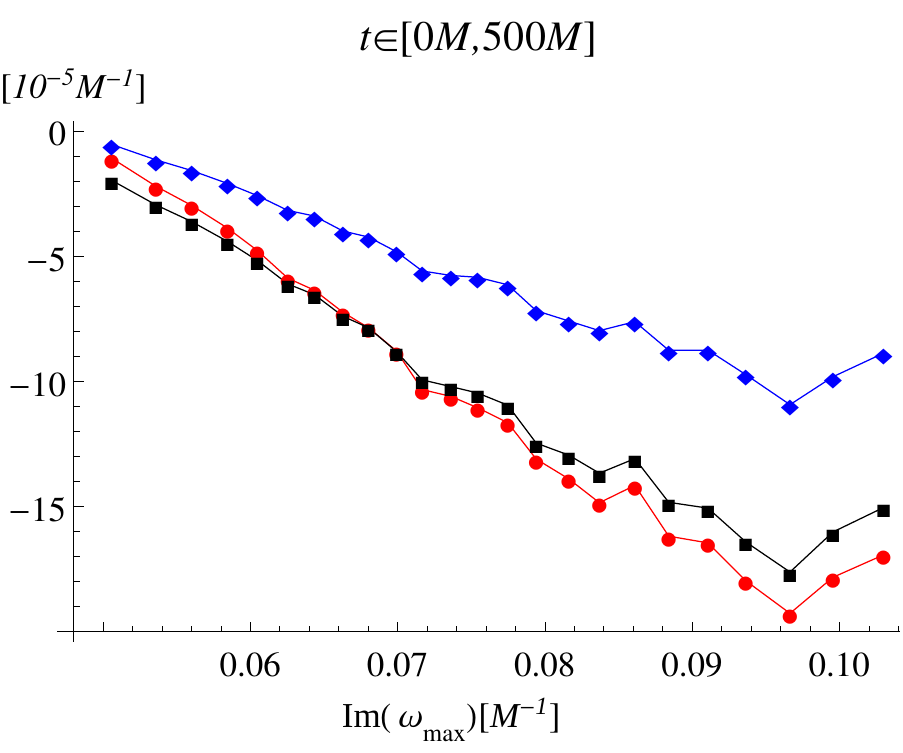}} \!\!\!
\raisebox{-0.5\height}{\includegraphics[width=0.44\textwidth]{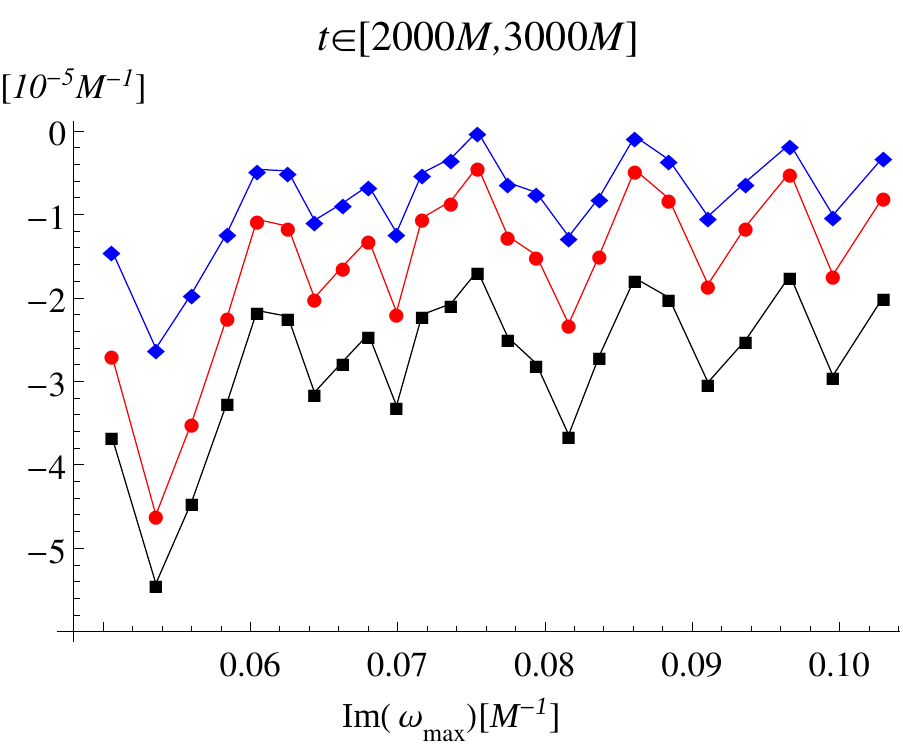}}\!
\raisebox{-0.5\height}{\includegraphics[width=0.115\textwidth]{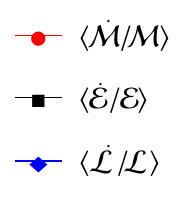}}
        \caption{Numerically obtained accretion rates of rest mass $\left\langle \dot{\mathcal{M}}/\mathcal{M}
         \right\rangle$, energy $\left\langle \dot{\mathcal{E}}/\mathcal{E} \right\rangle$ and angular momentum $\left\langle \dot{\mathcal{L}}/\mathcal{L}
         \right\rangle$  averaged over the evolution time intervals $[0,1000M]$ and $[2000M,3000M]$ as a function of the initial MRI growth rate estimate
         $\mathrm{Im}(\omega_{max})$. Note that a lower negative value of $\dot{\mathcal{M}},\dot{\mathcal{E}},\dot{\mathcal{L}}$ corresponds to a higher loss rate of the quantity from the torus and thus a higher accretion rate. The displayed values correspond to the parameter $\kappa$ in the 
        interval $[-1, 1.31]$.}
        \label{fig:ratecorel}
	\end{figure*}


\begin{figure}
    	\centering
		\includegraphics[width=0.49\textwidth]{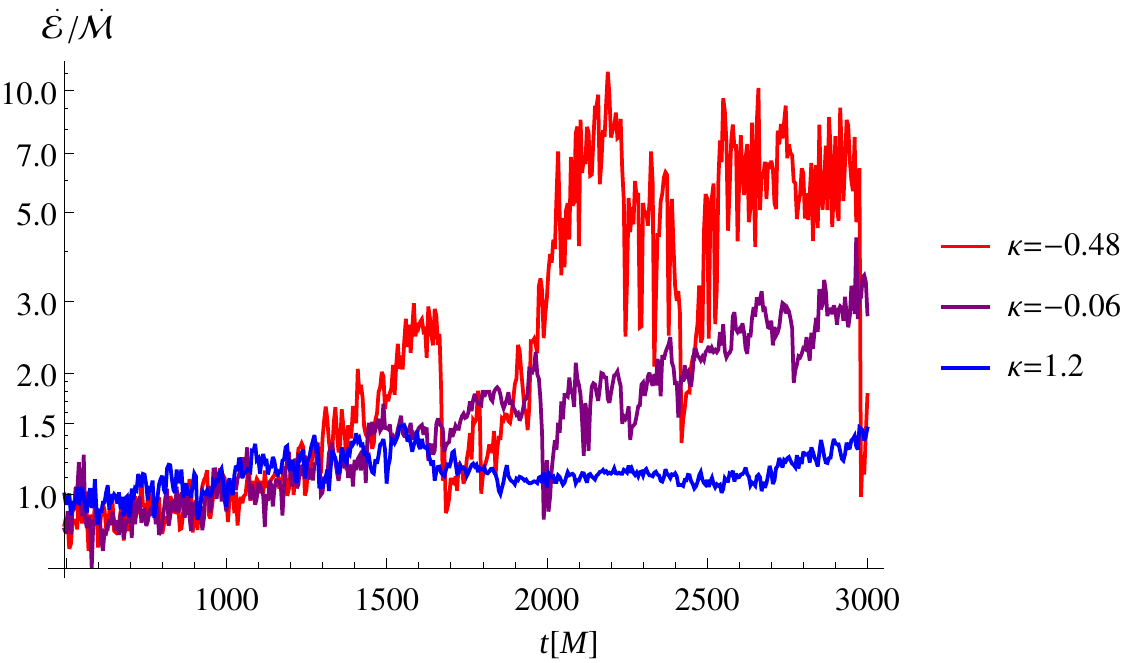}
        \caption{The accreted specific energy $\dot{\mathcal E}/\dot{\mathcal M}$ as a function of time for chosen values of the parameter $\kappa$.  (Color online.)} \label{fig:EdMd}
	\end{figure}



\begin{figure*}
    	\centering
		\includegraphics[width=0.2\textwidth]{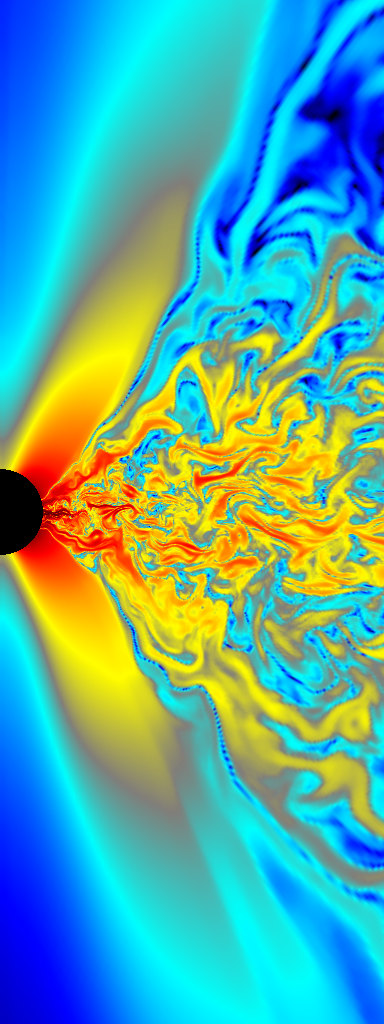}
        \includegraphics[width=0.2\textwidth]{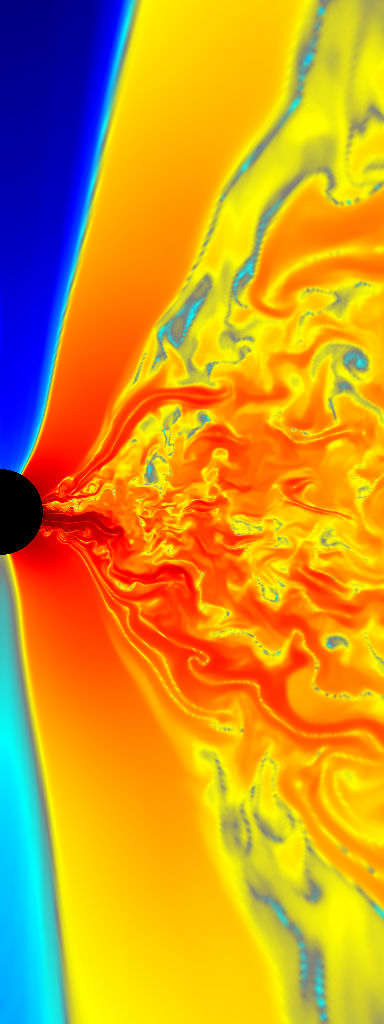}
        \includegraphics[width=0.2\textwidth]{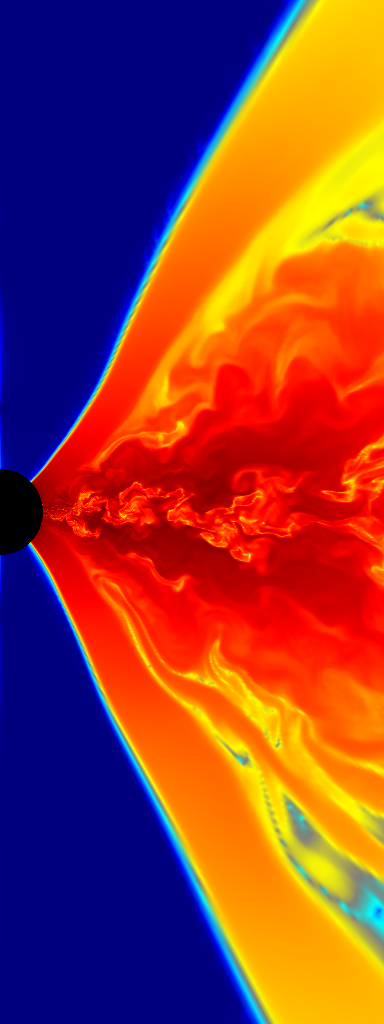}
        \includegraphics[width=0.2\textwidth]{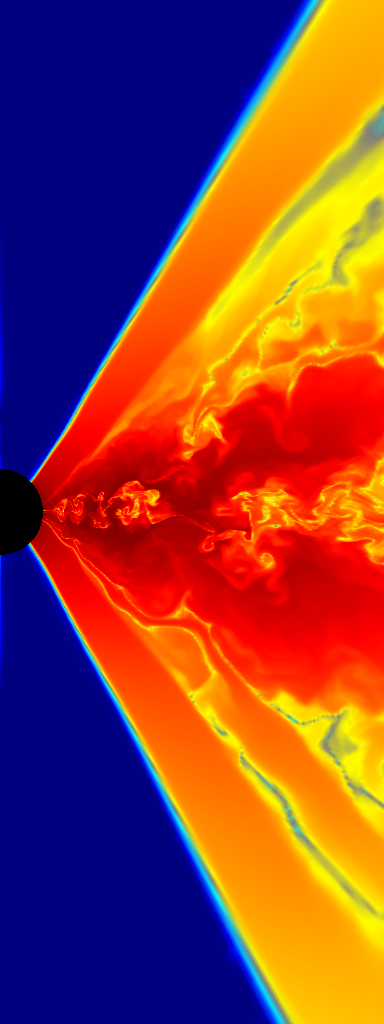}
        \caption{Snapshots of the structure of the magnetic fields for different tori evolutions plotted in Boyer-Lindquist coordinates. The color indicates $\log(b^2/b^2_\mathrm{max})$, where $b_\mathrm{max}^2$ is the maximal strength of the magnetic field in the respective image, the black circle is the interior of the black hole, and the right equatorial edge of the image is at $r=12M$. The first two images on the left are snapshots for the $\kappa=-0.48$ torus before and after the first $\dot{\mathcal{E}}/\dot{\mathcal{M}}$ peak, at $t=1200M$ and $t=1700M$ respectively (c.f. Fig. \ref{fig:EdMd}). The other two images on the right are snapshots for the ``$\dot{\mathcal{E}}/\dot{\mathcal{M}}$-quiescent'' $\kappa=1.2$ torus at $t=1200M$ and $t=1700M$ respectively. It appears that the behavior of $\dot{\mathcal{E}}/\dot{\mathcal{M}}$ is linked with the evolution of the coronal layer. (Color online.)} \label{fig:magsnaps}
	\end{figure*}



\begin{figure}
    	\centering
		\includegraphics[width=0.49\textwidth]{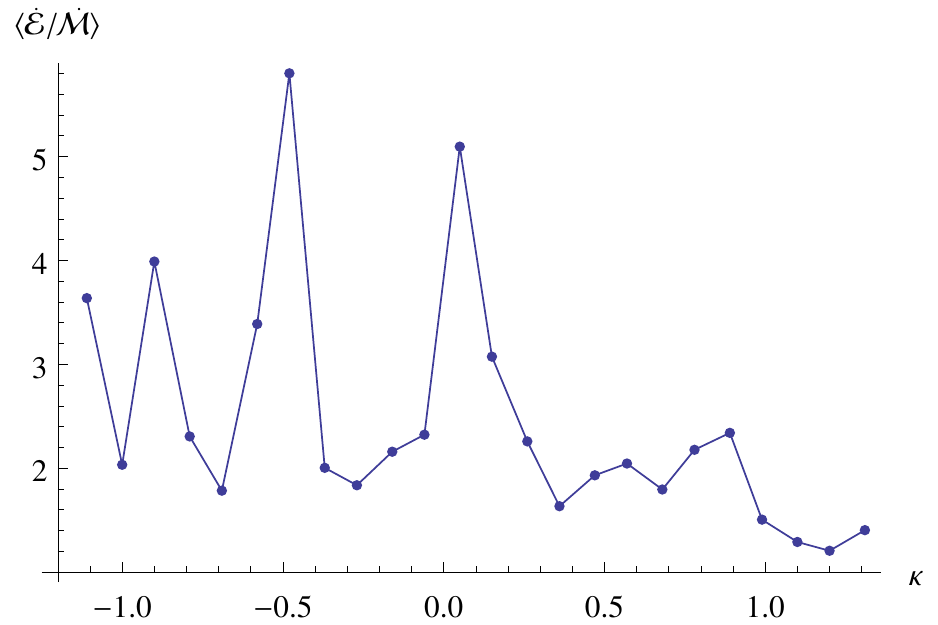}
        \caption{The accreted specific energy $\dot{\mathcal E}/\dot{\mathcal M}$ averaged over the interval $t \in [2000M,3000M]$ as a function of $\kappa$. The vertical axis has its origin at $1$, which means that all our flows return a negative nominal efficiency at late simulation times.} \label{fig:EdMdaverage}
	\end{figure}


   \begin{table} 
      \caption[]{Simulation parameters.}
\label{tab:parsim}
     $$ 
         \begin{array}{p{0.5\linewidth}l}
            \hline
            \noalign{\smallskip}
            Variable      &  \text{Value} \\
            \noalign{\smallskip}
            \hline
            \noalign{\smallskip}
            Black hole spin $a$ 		& 0.9375M     \\
            Resolution          & 512 \times 512 \\
            Inner radius of toroid     & 5M             \\
            Outer radius of toroid     & 12M             \\
            Horizon radius $r_\mathrm{H}$    & M+\sqrt{M^2-a^2}          \\
            Inner boundary of simulation    & 0.98 r_\mathrm{H}           \\
            Outer boundary of simulation     & 40M             \\           
            \noalign{\smallskip}
            \hline
         \end{array}
     $$ 
   \end{table}
%

   \begin{table}
\caption{Initial condition parameters. (Parameters  for tori with an isobaric contour passing through $r=5M,12M$ while $\lambda=0$.)}             
\label{tab:parinit}      
\centering                          
\begin{tabular}{c c | c c}        
\hline\hline                 
$\kappa$ & $\ell_0\,[M]$ & $\kappa$ & $\ell_0\,[M]$ \\    
\hline                        
-1.11	&	2.5889	&	0.15	&	3.2606	\\
-1.	&	2.6418	&	0.26	&	3.32871	\\
-0.9	&	2.69052	&	0.36	&	3.39239	\\
-0.79	&	2.74492	&	0.47	&	3.46449	\\
-0.69	&	2.79522	&	0.57	&	3.53203	\\
-0.58	&	2.85157	&	0.68	&	3.60865	\\
-0.48	&	2.90383	&	0.78	&	3.68053	\\
-0.37	&	2.96253	&	0.89	&	3.76223	\\
-0.27	&	3.0171	&	0.99	&	3.83902	\\
-0.16	&	3.07854	&	1.1	&	3.92644	\\
-0.06	&	3.13577	&	1.2	&	4.00877	\\
0.05	&	3.20034	&	1.31	&	4.10268	\\

\hline                                   
\end{tabular}
\end{table}

%
%
%

\section{Numerical simulations} \label{sec:sim}
Because the solutions presented in the last Section are in closed form, it is easy to set them up as initial conditions for the evolution in Kerr space-time in existing numerical codes and see whether they bring any variations in the  properties of the accretion disks that evolve from them. Hence, we implemented our initial conditions in the 2D HARM code of \citet{HARM,HARM2} {\bf and conducted the numerical study in a ``standard HARM set-up'', as described e.g. in \citet{mckinney2004}}. 

We placed the tori with various choices of the solution parameters near a Kerr black hole with spin $a = 0.9375M$, perturbed them with a small poloidal magnetic field with $P/b^2= 100$ at the pressure maximum of the torus, and evolved the tori for a time period of $3000M$ with a $512\times512$ resolution. {\bf As indicated already in \citet{balbus1991,balbus1998}, long-term MHD turbulence requires at least some magnetic field threading the disk vertically and this is the reason why the weak poloidal field is included in the initial conditions. Toroidal fields also trigger the MRI, but on their own they eventually decay \citep[see also][]{wielgus2015,fragile2017}. We have thus decided to not include a toroidal component of the magnetic field in the initial conditions ($\tilde{P}=0$) and leave the study of its effects to other works.}

What follows is a description of some of the properties of the chosen initial conditions and a discussion of the simulation results. We  used the default HARM outputs of energy, angular momentum, and rest-mass accretion rates to diagnose the evolutions. The parameters of the simulation are summarized in Table \ref{tab:parsim}.

\subsection{Properties of initial conditions} \label{subsec:properties}
Our main target was to investigate the properties of the tori with the angular-momentum relation (\ref{eq:ellstar}) in dependence on the various values of the parameter $\kappa$. We set $\lambda=0$ and determine $\ell_0$ by requiring that the toroids have their inner and outer edge at $r=5M$ and $r=12M$ in Boyer-Lindquist coordinates respectively. 

The thermodynamic relations were then determined by setting up the gas as initially isentropic, non-magnetized ($\tilde{P}=0$), and having a polytropic equation of state $P = K\mu^\gamma$, where $\mu$ is mass density and we set $\gamma = 13/9$ to roughly mimic an ideal-gas mixture of relativistic electrons and mostly non-relativistic ions. Furthermore, we used the HARM convention $K = \mu_\mathrm{c}^{1 - \gamma}$, where $\mu_c$ is the density of the toroid at the pressure maximum (see Appendix \ref{app:construction}). 

We constructed 24 such initial conditions with $\kappa$ in the interval $[-1.11,1.31]$; a summary of the resulting parameters is given in Table \ref{tab:parinit}, and the properties of the initial conditions can be inspected in Figures \ref{fig:Om} to \ref{fig:shapes}. 

We know from the analysis of von-Zeipel radii by \citet{abramowicz1978,kozlowski1978} that the $\ell=\mathrm{const.}$ toroids in the Kerr space-time will have an angular velocity along the equator falling off with radius, at least as long as we are above the photon orbit. Since $\kappa=0,\,\lambda=0$ corresponds to the $\ell=\ell_0=\mathrm{const.}$ tori, we can then assume that for moderate magnitudes of $\kappa$ our solutions will also have an angular velocity fall-off, at least if we are far away from the photon sphere. This expectation is fulfilled in the case of our tori, as can be seen on a few examples in Fig. \ref{fig:Om}.

Furthermore, by examining the angular momentum relation (\ref{eq:ellstar}) at $\lambda=0, \ell_0>0, \Omega>0$ and by assuming that angular velocity is falling off with radius, we conclude that $\kappa>0$ means angular momentum growing with radius, and a small $\kappa<0$ (such that the expression (\ref{eq:ellstar}) does not become singular) an angular momentum fall-off. This is demonstrated in Fig. \ref{fig:ell}.

{\bf As for the dependence of the vertical structure of the toroid on the parameter $\kappa$, that seems to be impossible to characterize by any simple analytical argument. We can only offer the observation that the angular momentum distribution is such that for $\kappa \gtrsim 1$ the rotation curve in the equatorial plane is closer to Keplerian values and thus the disk should be closer to the canonical ``thin disk'' also in the vertical structure. This can be seen to hold in Figs.  \ref{fig:z} and \ref{fig:shapes}. The $\kappa \gtrsim 1$ ($\lambda =0$) disks are thus the ones closest to the state assumed to arise from long-term MHD simulations and, as such, could be preferred as initial conditions in astrophysical studies.}

Last but not least, the HARM convention $K = \mu_\mathrm{c}^{1 - \gamma}$ leads to the specific enthalpy $h \equiv w/\mu$ to be equal to $h = (2 \gamma - 1)/(\gamma -1) = 4.25$ at the pressure maximum of every torus. This fact leads to the relativistic Bernoulli parameter $\mathcal{B} \equiv - h u_t$ being much larger than $1$ at the {\bf pressure maxima} of the tori, a property common to hot accretion flows \citep{narayan1994,narayan1995}. The values of the Bernoulli parameter at the pressure maxima of the tori as a function of $\kappa$ are plotted in Fig. \ref{fig:bernoulli}.

\subsection{Simulation results}
The general scenario of the evolution of any of our tori is {\bf up to some intermittent episodes (see Subsection \ref{subsec:effic}) qualitatively equivalent} to the default HARM simulation with the $\ell^* = \mathrm{const.}$ tori as described e.g. in \citet{mckinney2004}. We see the onset of the MRI and a rapid transition to turbulence throughout the disk starting from the inner edge. In parallel, the inner edge extends all the way towards the black hole, forming a plunging region, and the accretion disk reaches a quasi-steady state with a magnetically dominated corona layer above it, and a nearly evacuated magnetized funnel around the poles. 

The simulation time $t = 3000 M$ is about 10 orbital periods of the pressure maxima and it is sufficient to see the establishment of these quasi-stationary structures. However, it is not enough to see the ultimate fate of the disk in terms of the accumulation of magnetic flux around the black hole and a possible transition to a magnetically arrested disk, or the spreading of the disk to large radii through angular momentum exchange \citep[for a long simulation see][]{narayan2012}. 

{\bf The mathematical analysis of \citet{balbus1991} showed that the MRI grows with a rate which is dependent both on the wave vector of the disturbance and the local angular-velocity gradients in the fluid. To obtain a single numerical estimate of some kind of ``typical strength'' of the MRI for each torus, we have chosen to use the growth rate of the fastest-growing MRI mode (maximum of the MRI growth rate in the perturbation wave-vector space) $\mathrm{Im}(\omega_\mathrm{max})$ evaluated at the location of the pressure maxima of the tori.} \citet{gammie2004} showed that the rate of the fastest growing mode can be computed as half the shear rate in the case of relativistic Keplerian disks, $-\omega^2_\mathrm{max} = \sigma^2/2$ and, even though this result has not been rigorously proven to apply to general relativistic flows, we use it here as our estimate \citep[for definition and computation of shear rates in Kerr space-times see e.g.][]{semerakcirc}. 

The resulting MRI growth-rate estimates are given in Fig. \ref{fig:MRIestim}, where we see that by variation of $\kappa$ we get an MRI e-folding time from $10M$ to about $20M$. As can be seen on a few selected examples of accretion rates in Fig. \ref{fig:accexamples}, this leads to the first wave of matter arriving at the black hole at around $100M-200M$. This certainly does not mean that the MRI needs $\sim 10$ e-folds to take effect because this time also includes the inspiral time required for the matter released by the instability to arrive to the black hole horizon. In fact, the waves arrive within a few tens of $M$ away from each other, which suggests that matter is released from the tori withing $\sim 5$ e-folding times of the MRI.

This observation is further documented in Fig. \ref{fig:ratecorel}, where we see that the average accretion rates over the first $500M$ of evolution are correlated with the initial MRI estimates. This may be explained by noticing from Figure \ref{fig:accexamples} that the ``first accretion waves'' arrive earlier from the tori with higher shear (smaller $\kappa$), but also that these waves bring larger amounts of matter with them. However, we also see that the correlation starts to be broken for the most shearing tori (with largest MRI growth rates), which is probably because they are already entering the non-linear mode of evolution in the chosen averaging interval. 

Furthermore, if we try to see whether the accretion rates in the last $1000M$ of our evolution are correlated to the initial $\mathrm{Im}(\omega_\mathrm{max})$, we see in the second part of Fig. \ref{fig:ratecorel} that there is no tangible correlation. This is because by that time the tori have evolved to the quasi-stationary ``asymptotic\footnote{Meaning that they then evolve on timescales much longer than the orbital time}'' states where their relation to the properties of initial conditions has mostly been washed out. (Perhaps up to the energy and angular-momentum content, see next Subsection.)

\subsection{Advected specific energy and nominal efficiency} \label{subsec:effic}

We also computed the ``nominal efficiency'' $\tilde{\eta} \equiv 1 - \langle \dot{\mathcal E}/\dot{\mathcal{M}} \rangle$ as was done in \citet{mckinney2004} but we found negative efficiencies for all of our tori at late times. {\bf This is not a mistake but, as will be seen from the following, a robust property of this type of simulations with an analytical underpinning. Hence, we must warn against the usage of $\tilde{\eta}$ as any sort of efficiency of the accretion process. Let us now discuss the arguments which explain this phenomenon.}

The total energy in the disk is defined as
\begin{equation}
\mathcal{E}(t) = -\int_{V} T^{t}_{\;t} \sqrt{-g} \,\d^3 x = \int \mathcal{B} \mu u^t \sqrt{-g}  - P\sqrt{-g} \,\d^3 x + \mathcal{E}_\mathrm{m}\,,
\end{equation}
where $V$ is the spatial volume, and $\mathcal{E}_\mathrm{m}$ is the energy in the magnetic fields (initially almost zero). The total rest mass $\mathcal{M}$ is, on the other hand, defined as
\begin{equation}
\mathcal{M}(t) = \int_V \mu u^t \sqrt{-g} \,\d^3 x\,.
\end{equation}
The expression $\mu u^t \sqrt{-g}$ is just the coordinate mass density, so for instance $\int \mathcal{B} \mu u^t \sqrt{-g} \d^3 x /\int \mu u^t \sqrt{-g} \d^3 x $ is simply the mass-averaged Bernoulli parameter. 

Thus, we can see that the ratio $\mathcal{E}/\mathcal{M}$ corresponds to the average Bernoulli parameter with contributions to energy from the internal stress in the gas, and the magnetic-field energy. A similar argument leads us to the realization that $\dot{\mathcal E}/\dot{\mathcal M}$ is the average Bernoulli parameter of the accreted fluid elements plus similar terms. However, since we start with a toroid with most of its elements having $\mathcal{B} >1$ (see Fig.  \ref{fig:bernoulli}), we should not strictly expect the advection dominated flow to accrete exclusively the portion of the matter with $\mathcal{B}<1$ and spit the rest out in an outflow. Thus, even when ignoring the pressure and magnetic contribution to energy, we should not be surprised by $\dot{\mathcal E}/\dot{\mathcal M}>1$.

The simulation of \citet{mckinney2004} did not find the average $\langle \dot{\mathcal E}/\dot{\mathcal M}\rangle$ to be larger than $1$ but we believe that this is only thanks to the fact that their simulation halted at $t=2000M$. In contrast, we found that for times $t>2000M$ the irregular advection mechanism is able to push even the high-energy fluid elements into the black hole. This is demonstrated in Figure \ref{fig:EdMd}, where we see that $\dot{\mathcal E}/\dot{\mathcal M}$ undergoes erratic ``build-ups'' whose origin seems to be linked with the behavior of the magnetic fields and the coronal layer (see Fig. \ref{fig:magsnaps}). 

Specifically, visual inspection of the plots of the strength of magnetic fields reveals that the build-ups in $\langle \dot{\mathcal E}/\dot{\mathcal M}\rangle$ are associated with the coronal layer slowly raising all the way to the axis and filling the polar funnel; the abrupt drop in accreted specific energy is then associated with a sudden evacuation of the axis region and the reemergence of the funnel. This of course suggests that what is accreted in the ``super-energetic'' periods is mostly the magnetically dominated matter from the coronal region around the poles. On the other hand, it also suggests that this phenomenon might be associated with the strictly axi-symmetric structure of the simulation and that it would not be reproduced in a 3D simulation. Alternatively, it could be caused by the small topological defect which is introduced in HARM to regularize the coordinate singularity at the axis \citep[see][]{HARM}. 

One thing which is clear to us is the fact that this ``super-energetic'' accretion cannot be easily related to any parameter of the initial conditions, as we show in Fig. \ref{fig:EdMdaverage}. All the properties of the initial conditions vary as smooth, typically monotonous functions of the parameter $\kappa$, so if the late $\langle \dot{\mathcal E}/ \dot{\mathcal M} \rangle$ is to depend on any simple property of the initial conditions, it should be obvious from its dependence on $\kappa$.  Even though the late average $\langle \dot{\mathcal E}/ \dot{\mathcal M} \rangle$ can be recognized in Fig. \ref{fig:EdMdaverage} as a continuous function of $\kappa$ in certain regions, the sampling is not sufficient to understand the structure any further. {\bf The only other observation, inferred from Figs. \ref{fig:ell} and \ref{fig:EdMdaverage}, is that the extremely high values of specific energy seem to be associated with initial rotation curves which are very far away from a Keplerian profile.}

We believe that this dependence of $\langle \dot{\mathcal E}/ \dot{\mathcal M} \rangle$ on initial conditions should be studied in more detail in future works. Additionally, a more careful study should separate $ \dot{\mathcal E}/ \dot{\mathcal M}$ from the actually accreted Bernoulli parameter because the former involves a significant, or even dominant contribution of the energy of the magnetic field at late evolution times. However, implementing and running diagnostics that would allow to make such a distinction and shed more light on this numerically observed phenomenon is out of the scope of the current paper.

\section{Conclusions}

   \begin{enumerate}
      \item We presented generalizations of the known closed-form analytical solutions of geometrically thick fluid tori near black holes; our solutions are easy to construct and implement in numerical codes. Unlike the previously known solution families, our family provides a rich spectrum of possibilities for the rotation curve and geometrical shapes even if parameters such as the inner and outer radii are constrained. {\bf In particular, our $\kappa \gtrsim 1$ disks have semi-Keplerian rotation profiles and could thus be preferred for initial conditions of astrophysical simulations rather than the $\ell=\mathrm{const.}$ \citep{abramowicz1978,kozlowski1978} or $\ell^* = \mathrm{const.}$ \citep{fish-mon} tori.}
      \item The MRI growth rates for different tori from our new class have different magnitudes and the tori exhibit different numerical evolutions accordingly. Namely, the average normalized accretion rate $\langle \dot{\mathcal{M}}/\mathcal{M} \rangle$ over the first $500M$ of the simulation ($\lesssim$ 3 orbital periods) turns out to be linearly proportional {\bf to the growth rate of the fastest-growing MRI mode at the pressure maximum of the torus}.
      \item On the other hand, the asymptotic states of the accretion disk which are achieved for an evolution time $\gtrsim 2000 M$ do not exhibit any correlation of accretion rates with the initial MRI growth rates. Additionally, all of the asymptotic accretion disks tend to have the same qualitative disk-corona-funnel structure as described by e.g. \citet{mckinney2004} and the normalized accretion rates of their angular momentum, energy and mass vary within a single order of magnitude. 
      \item Nevertheless, we see that the disks vary quite wildly in the specific energy of the fluid they accrete; for particular choices of initial conditions and at the most extreme periods, the energy of the accreted elements approaches ten times their rest mass. {\bf This seems to be connected with the behavior of the strongly magnetized coronal layer because the extreme accretion episodes are associated with the corona pushing into the funnel, sometimes even up to the point of the short disappearance of the evacuated region. This behavior, however, appeared only for initial conditions far away from a Keplerian rotation profile.} Further insights into this question should be obtained by future studies which should also involve more realistic matter models for the disk and better diagnostics ran during the simulation.
   \end{enumerate}

\begin{acknowledgements}
      We would like to thank Claus Lämmerzahl and Volker Perlick for supervision and useful discussions on the topic. We are grateful for the support from a Ph.D. grant of the
German Research Foundation (DFG) within Research Training
Group 1620
``Models of Gravity''. Additionally, PJ kindly acknowledges the support from the Erasmus Mundus Joint Doctorate IRAP program.

\end{acknowledgements}


\begin{appendix}
\section{Deriving various forms of Euler equation} \label{app}
\subsection{Simplifying $\left(T^{\mu}_{\;\; \nu\mathrm{(m)}}\right)_{;\mu}$}
Omitting terms which are zero due to the stationarity and axisymmetry of all quantities, the circularity of velocity field, and toroidality of magnetic field, we obtain
\begin{equation}
\left(T^{\mu}_{\;\; \nu\mathrm{(m)}}\right)_{;\mu} = b^2 a_\nu + \frac{(b^2)_{,\nu}}{2} - b_{\nu;\mu} b^\mu\,. 
\end{equation}
Let us now rewrite the $a_\nu$ and $b_{\nu;\mu}b^\mu$ terms as
\begin{align}
u_{\nu;\mu} u^\mu = \frac{1}{2} g^{\alpha \beta}_{\;\;\;,\nu} u_\alpha u_\beta \,,\\
b_{\nu;\mu} b^\mu = \frac{1}{2} g^{\alpha \beta}_{\;\;\;,\nu} b_\alpha b_\beta \,.
\end{align}
We then obtain
\begin{align}
&\left(T^{\mu}_{\;\; \nu\mathrm{(m)}}\right)_{;\mu} = -\frac{b^2}{2}g^{\alpha \beta}_{\;\;\;,\nu} \left(-u_\alpha u_\beta + \frac{b_\alpha b_\beta}{b^2} \right)   + \frac{(b^2)_{, \nu}}{2} \,.
\end{align}
Now, we realize that $u_\mu$ and $b_\mu/\sqrt{b^2}$ are two normalized orthogonal vectors exclusively in the $t-\varphi$ direction and we thus have
\begin{equation}
\frac{u_\alpha u_\beta}{u_\mu u^\mu} + \frac{b_\alpha b_\beta}{b^\mu b_\mu} = - u_\alpha u_\beta + \frac{b_\alpha b_\beta}{b^2} = g_{\alpha \beta}|_{(\varphi t)}\,,
\end{equation} 
where $g_{\alpha \beta}|_{(\varphi t)}$ is the $\varphi,t$-restriction of the metric (that is, $g_{rr}|_{(\varphi t)}= g_{\theta \theta}|_{(\varphi t)} = 0$ but otherwise the same as the metric). We then see that the magnetic part of the structural equations reads
\begin{align}
&\left(T^{\mu}_{\;\; \nu\mathrm{(m)}}\right)_{;\mu} = -\frac{b^2}{2}g^{\alpha \beta}_{\;\;\;,\nu} g_{\alpha \beta}|_{(\varphi t)}  + \frac{(b^2)_{,\nu}}{2} \,.
\end{align}
We now compute 
\begin{equation}
\begin{split}
&g^{\alpha \beta}_{\;\;\;,\nu} g_{\alpha \beta}|_{(\varphi t)} = (g^{\alpha \beta}|_{(\varphi t)})_{,\nu}\, g_{\alpha \beta}|_{(\varphi t)} = - g^{\alpha \beta}|_{(\varphi t)} (g_{\alpha \beta}|_{(\varphi t)})_{,\nu}  
\\&= -\mathrm{Tr}\left[\mathbf{g}|_{(\varphi t)}^{\!-1} (\mathbf{g}|_{(\varphi t)})_{,\nu} \right] =  -\frac{\rho^2_{,\nu}}{\rho^2}\,, \label{eq:detform}
\end{split}
\end{equation}
where in the last step we have used the formula for the derivative of a determinant of a matrix, and we denote $\rho^2 \equiv- \mathrm{Det}\left(\mathbf{g}|_{(\varphi t)}\right)= g_{t\varphi}^2- g_{tt}g_{\varphi \varphi}$ . Using (\ref{eq:detform}) we then obtain that the magnetic part of the stress-energy tensor can be written as
\begin{align}
&(T^{\mu \mathrm{(m)}}_{\; \,\nu})_{;\mu} = \frac{(\rho^2 b^2)_{,\nu}}{2 \rho^2 } \,.
\end{align}


\subsection{Simplifying $\left(T^{\mu}_{\;\; \nu\mathrm{(f)}}\right)_{;\mu}$}
Under the symmetry assumptions and circularity of the flow we obtain
\begin{equation}
\left(T^{\mu}_{\;\; \nu\mathrm{(f)}}\right)_{;\mu} = w a_\nu + P_{,\nu}\,. 
\end{equation}
Let us now briefly derive the two possible forms of $a_\nu$ useful for the construction of analytical solutions.


\subsubsection{Fishbone-Moncrief form of acceleration}
The Fishbone-Moncrief form is obtained by expressing the acceleration in terms of the velocity in the ZAMO frame. The linear velocity of the flow in the ZAMO frame is $V =  u_\varphi/\sqrt{g_{\varphi\varphi}}$, and the four-velocity components of the circular flow are given in terms of it as
\begin{align}
&u_\varphi = \sqrt{g_{\varphi \varphi}} V\,,\\
&u_t = -\sqrt{\frac{1 + V^2}{-g^{tt}}} - \omega \sqrt{g_{\varphi \varphi}} V\,,
\end{align} 
These expressions are then directly substituted into the four-acceleration $a_\nu = g^{\alpha \beta}_{\;\;\;,\nu} u_\alpha u_\beta/2$ to obtain
\begin{align}
\begin{split}
a_\nu
&= \frac{1}{2} \left( g^{tt}_{\;\;\;,\nu} u_t^2 + 2 g^{t\varphi}_{\;\;\;,\nu} u_t u_\varphi + g^{\varphi \varphi}_{\;\;\;,\nu} u_\varphi^2 \right)
\\
&=-\ln(\mathcal{R})_{,\nu} V^2 + \omega_{,\nu} \mathcal{R} V\sqrt{1 + V^2} + \Phi_{\mathrm{(Z)},\nu}\,.
\end{split}
\end{align}


\subsubsection{``Polish-donut'' form of acceleration}
The Polish-donut form of acceleration is derived by realizing that the acceleration can be rewritten as
\begin{equation}
\begin{split}
& u_{\nu;\mu} u^\mu = \frac{1}{2} g^{\alpha \beta}_{\;\;\;,\nu} u_\alpha u_\beta = \frac{1}{2} \left[(g^{\alpha \beta} u_\alpha u_\beta)_{,\nu} +2 u_{\alpha} u^\alpha_{\;,\nu} \right] 
\\& = u_{\alpha} u ^\alpha_{\;,\nu} =  u_{t} u^t_{,\nu} + u_\varphi u^\varphi_{\;,\nu}\,. \label{eq:PDacc1}
\end{split}
\end{equation}
We now use the four-velocity normalization to express $u_t = -(1+u^\varphi u_\varphi)/u^t$ and obtain 
\begin{equation}
\begin{split}
& a_\nu = -\frac{u^t_{,\nu}}{u^t} + \frac{u^t_{,\nu}}{u^t} u^\varphi u_\varphi + u_\varphi u^\varphi_{\;,\nu} = - \ln(u^t)_{,\nu} + u_\varphi u^t \left(\frac{u^\varphi}{u^t} \right)_{,\nu}\,. \label{eq:PDacc2}
\end{split}
\end{equation}
The last step is to use the four-velocity normalization to obtain the identity $u^t u_t = -1/(1 - \Omega \ell)$ and thus $u_\varphi u^t = \ell/(1 - \Omega \ell)$ which leads to equation (\ref{eq:pdform}).


\section{Construction of tori} \label{app:construction}
When we postulate $\ell = (\ell_0+ \lambda \Omega)/(1+\kappa \ell_0 \Omega)$, we obtain the angular rotation frequency for $\kappa \neq 1$ as
\begin{equation}
\begin{split}
&\Omega = \frac{ \mathcal{A} - \sqrt{\mathcal{A}^2 + 4 (\ell_0 g^{\varphi\varphi}  - g^{t\varphi})( \lambda g^{t\varphi} - \kappa \ell_0 g^{tt}) }}{2 ( \lambda g^{t\varphi} - \kappa \ell_0 g^{tt})} \\
&\mathcal{A} =  g^{tt} + \lambda g^{\varphi\varphi} - \ell_0 g^{t\varphi}(1+ \kappa) \label{eq:Omnase}
\end{split}
\end{equation}
{\bf When we have $\ell = \ell_0$ (such as when $\kappa=\lambda=0$), the formerly quadratic equation for $\Omega$ in (\ref{eq:Omkvadra}) becomes linear and the quadratic root (\ref{eq:Omnase}) has a formal singularity. In that case we can either take an appropriate limit or directly substitute into equation (\ref{eq:Omzlomek}) to obtain}
\begin{equation}
\Omega = \frac{\ell_0 g^{\varphi \varphi} - g^{t \varphi}}{\ell_0 g^{t\varphi} - g^{tt}}\,.
\end{equation}
The function $L(\Omega)$ from (\ref{BigL}) is easily integrated as (for a general choice of parameters)
\begin{align}
\begin{split}
 L = &-\frac{1}{2}\ln\left[1+(\kappa-1) \ell_0 \Omega - \lambda \Omega^2\right] 
\\  & - \frac{\ell_0 (1 + \kappa)}{2\mathcal{C}} \ln \left[ \frac{\mathcal{C} - \ell_0(1-\kappa) - 2 \lambda \Omega}{\mathcal{C}+ \ell_0(1-\kappa) + 2 \lambda \Omega} \right] \,, 
\\
\mathcal{C} = &\sqrt{\ell_0^2(\kappa-1)^2 + 4 \lambda} \,.
\end{split}
\end{align}
There are degeneracies in the parametrization when we allow for $\lambda\neq 0$ and we thus recommend to set $\lambda=0$ for most practical purposes. Then the function $L$ simplifies as
\begin{equation}
L = \frac{1}{\kappa-1}\ln\left[1+(\kappa-1) \ell_0 \Omega\right] \,.
\end{equation}
The final expression for the thermodynamic potentials (for $\lambda=0$) then reads
\begin{align}
\begin{split}
W + \tilde{W} =&-\ln\left[ \left[1 + (\kappa-1) \ell_0 \Omega\right]^{1/(\kappa-1)}\right] \\& -\ln \left[\sqrt{-g_{tt} - 2 g_{t \varphi} \Omega - g_{\varphi \varphi} \Omega^2 } \right]\,, \label{eq:Wnase}
\end{split}
\end{align}
where we of course need to substitute the $\lambda=0$ version of (\ref{eq:Omnase}) to obtain a purely coordinate-dependent expression.

The expressions for mass and enthalpy density of an isentropic, non-magnetized, and polytropic fluid $P = K \mu^\gamma$ read \citep[c.f.][]{rezzola}
\begin{align}
\label{eq:Density}
&\mu = \left[\frac{\gamma-1}{K\gamma}\Big(\exp [W-W_\mathrm{s}] - 1\Big)\right]^{1/(\gamma-1)} \,,\\
&w = \mu \exp [W-W_\mathrm{s}] = \mu + \frac{K \gamma}{\gamma -1} \mu^{\gamma}\,.
\label{eq:SpecEnth}
\end{align}
where $W_\mathrm{s}$ is the value of $W$ at the surface of the torus (conventionally the inner edge).

The dynamical properties of the polytropic gas are invariant with respect to $K$ \citep[see e.g.][]{rezzola}, so the choice of $K$ is somewhat arbitrary. We choose $K$ the same way as in the original HARM code and that is so that
\begin{align}
&\frac{\mu}{\mu_\mathrm{c}} = \left[\frac{\gamma-1}{\gamma}\Big(\exp [W-W_\mathrm{s}] - 1\Big)\right]^{1/(\gamma-1)} \,,
\end{align}
where $\mu_\mathrm{c}$ is the maximum density\footnote{In the HARM code, one actually stores and evolves the dimensionless $\mu/\mu_\mathrm{c}$ instead of $\mu$.} in the torus (i.e. at the pressure maximum). This leads immediately to $K = \mu_\mathrm{c}^{1-\gamma}$. 

However, as was mentioned in the main text and can be seen from (\ref{eq:Density}) and (\ref{eq:SpecEnth}), the choice of $K$ {\em does influence} the absolute values of specific enthalpy $h=w/\mu$ throughout the fluid. Hence, some of the important physical characteristics of the toroids such as their total specific energy $\mathcal{E}/\mathcal{M}$ or total specific angular momentum $\mathcal{L}/\mathcal{M}$ depend on the value of $K$ even at fixed spatial extent and rotation curves!

If we then fix $\kappa$ (and $\lambda=0$), and we want to confine the torus between some $r_\mathrm{in}$ and $r_\mathrm{out}$, we must numerically solve the following equation for $\ell_0$ using either well-known algorithms or software such as Mathematica or Matlab
$$W(r = r_\mathrm{in},\theta =\pi/2;\kappa,\ell_0) - W(r = r_\mathrm{out},\theta =\pi/2;\kappa,\ell_0) = 0\,.$$
The parameters which we obtained in the case of $r_\mathrm{in} = 5M,\, r_\mathrm{out}=12M$ are given in Table \ref{tab:parinit}.

\end{appendix}

\bibliographystyle{aa}
\bibliography{literature}

\end{document}